\documentclass[12pt]{article}
\usepackage{amssymb,amsmath,graphicx}\pdfoutput=1
\begin{document}
\title{\textbf{From confinement to dark energy}} 
\author{B.~Holdom%
\thanks{bob.holdom@utoronto.ca}\\
\emph{\small Department of Physics, University of Toronto}\\[-1ex]
\emph{\small Toronto ON Canada M5S1A7}}
\date{}
\maketitle
\begin{abstract}
The infrared divergence of the self-energy of a color charge is due to an enhancement of the long wavelength modes of the color Coulomb potential field. There are also long wavelength contributions to the QCD vacuum energy that are similarly enhanced. Vacuum modes of Hubble scale wavelengths may be affected in a cosmological setting and this can lead to a residual positive energy density of the form $H^d\Lambda_{\rm QCD}^{4-d}$. Lattice studies constrain $d$. If the dark energy takes this form then the universe is driven towards de Sitter expansion, and we briefly study this cosmology when $d$ is just slightly above unity.
\end{abstract}

\section{Introduction}
There have been recent suggestions \cite{s1,s2,s3} that in an expanding universe with Hubble parameter $H$, QCD can produce a dark energy density of order $H\Lambda_{\rm QCD}^3$. This is an intriguing claim since numerically this energy density is quite close to (although somewhat above) the observed value, and it leads to a cosmology not that dissimilar to the cosmological constant cosmology. But there are reasons to be skeptical. Given that QCD has a mass gap there would seem to be a clear decoupling between the vastly different QCD and Hubble length scales. One can use the Casimir effect as an analogy, where the effects of massive fields are exponentially suppressed. And even with a massless field the expected contribution to vacuum energy is only of order $H^4$, again in analogy with the Casimir effect.

For a result different from the Casimir effect, one can consider an infinite volume and simply remove the contribution of the large wavelength modes to the zero-point vacuum energy. Then the change in vacuum energy density is
\begin{equation}
-\int^H \frac{d^3k}{(2\pi)^3}\frac{1}{2}\omega_k\sim -H^3m
,\label{e13}\end{equation}
where we give the example of a massive field with $m\gg H$. This could be considered the maximum effect that Hubble scale physics could have on a Hubble scale contribution to vacuum energy. But this is still a miniscule contribution and it is of the wrong sign. The $H\Lambda_{\rm QCD}^3$ behavior is said to rely on various properties of QCD, including strong interactions, topological effects and ghosts that decouple in flat spacetime. But it is presently unclear to us how the arguments of  \cite{s1,s2,s3} and the related work in lower dimensions \cite{s6,s7} overcome the naive expectations.

We shall define $\Delta\rho_{\rm vac}$ to be the $H$ dependent contribution to the QCD vacuum energy density due to effects on Hubble scales. $\Delta\rho_{\rm vac}$ should be finite and its dependence on $H$ means that it can be discussed independently of the cosmological constant problem. From the previous remarks, if $\Delta\rho_{\rm vac}$ is to be large enough to be of any interest there must be some enhancement of the long wavelength contributions to the QCD vacuum energy. We shall argue that existing lattice studies can help to establish whether or not such an enhancement exists.

We find particularly relevant a set of studies that are helping to elucidate the mechanism of confinement. The Coulomb gauge picture of confinement, originally due to Gribov \cite{s5} and developed over the years \cite{s15}, connects confinement to the nontrivial structure of the gauge field configuration space (as manifested by Gribov copies). It has received support from numerous lattice studies, with \cite{s4} being a recent example. Intrinsic to this picture of confinement is the nontrivial scaling behavior of various quantities in the far infrared. It is this type of effect that is of interest to $\Delta\rho_{\rm vac}$.

On the lattice some finite volume analog of $\Delta\rho_{\rm vac}$ could in principle be measured without choosing a gauge. But this would be difficult in practice and so more useful is the fact that the lattice can provide information on the scaling behavior of 2-point functions in a particular gauge. This information can then be carried over to a continuum description where a gauge choice is necessary, and where a direct estimate of $\Delta\rho_{\rm vac}$ (a physical quantity) is more feasible.\footnote{The use of gauge-fixed lattice results in this way has recently been extolled in \cite{s16}.} The choice of the Coulomb gauge is also appropriate for another reason, its non-covariance matches the existence of a preferred frame in the cosmological setting.

What is of particular interest in Coulomb gauge is the infrared behavior of the longitudinal component of the color electric field, the color Coulomb potential. The massless long wavelength modes of this field are enhanced, and this enhancement must be sufficiently large to be consistent with confinement. We shall follow a treatment \cite{s8} that derives a necessary condition for confinement in terms of the infrared divergence of the Coulomb self-energy of a color charge. In contrast the propagation of the transverse modes of the gauge field are suppressed in the infrared. This behavior is also related to confinement and it can be described by an effective mass function $m(k)\approx\Lambda_{\rm QCD}^2/k$ for these modes.\footnote{It was pointed out in \cite{s2} that this would enhance the contribution in  (\ref{e13}), but not sufficiently and without correcting the sign.}

We then turn to the vacuum energy and explore an interplay between the nonperturbatively enhanced long wavelength longitudinal modes and the modes that have developed a mass gap. The result is an infrared enhancement of $\Delta\rho_{\rm vac}$. We end section 2 by making use of a recent lattice study \cite{s9} to help determine the enhancement. In section 3 we return to the cosmological setting and discuss how $\Delta\rho_{\rm vac}$ feeds back and affects cosmological evolution. An interesting cosmology arises.

$\Delta\rho_{\rm vac}$ depends on $H$ by definition and implicit in our picture is that there is no $H$ independent contribution to the vacuum energy; only then is the Minkowski vacuum consistent. Why this should be the case is the original cosmological constant problem. We only remark here on what we feel to be a necessary condition for a solution: the ultimate theory should contain no explicit dimensionful parameters and all masses should arise dynamically. The quantity $\Delta\rho_{\rm vac}$ of interest here is an example of a vacuum energy that appears due to the introduction of the dimensionful quantity $H$.

\section{The enhancement}
In Coulomb gauge where $\nabla_i A_i^a=0$, the color electric field of QCD can be written as $E_i^a-\nabla_i\varphi^a$ where $\nabla_i E_i^a=0$. Gauss's law determines the color Coulomb potential $\varphi^a$ in terms of the transverse gluons and quarks.
\begin{equation}
\int d^3y M_{xy}^{ab}\varphi^b(y)=\rho^a(x)
\end{equation}
\begin{equation}
M^{ab}_{xy}=(-\delta^{ab}\nabla^2+g f^{abc}A_i^c(x)\nabla_i)\delta(x-y)
\end{equation}
\begin{equation}
\rho^a=gq^\dagger T^a q-gf^{abc}A^b_i E^c_i
\label{e6}\end{equation}
The Coulomb energy is
\begin{eqnarray}
H_{\rm coul}&=&-\frac{1}{2}\int d^3y\langle\varphi^a(y)\nabla^2\varphi^a(y)\rangle\\
&=&\frac{1}{2}\int d^3xd^3yd^3z \langle\rho^a(x)[M^{-1}]_{xy}^{ab}[-\nabla^2_y][M^{-1}]_{yz}^{bc}\rho^c(z)\rangle
\end{eqnarray}
$M$ is the Fadeev-Popov operator and we consider its eigenfunctions in an arbitrarily large volume,
\begin{equation}
\int d^3y M_{xy}^{ab}\phi_n^b(y)=\lambda_n\phi_n^a(x)
,\end{equation}
normalized such that
\begin{equation}
\int d^3y \phi^{*a}_m(y)\phi^a_n(y)=\delta_{mn}
.\end{equation}
Then we have the representation
\begin{equation}
[M^{-1}]_{xy}^{ab}=\sum_n \frac{\phi_n^a(x)\phi_n^{*b}(y)}{\lambda_n}
\end{equation}
which gives
\begin{equation}
H_{\rm coul}=\frac{1}{2}\sum_{mn} \frac{F_{mn}}{\lambda_m\lambda_n}\int d^3xd^3y\phi_m^a(x)\langle\rho^a(x)\rho^c(z)\rangle\phi_n^{*c}(z)
\label{e1}\end{equation}
where
\begin{equation}
F_{mn}=\int d^3x\phi_m^{*a}(x) (-\nabla^2)\phi_n^a(x)
.\end{equation}
This is a useful formulation since the eigenfunctions and eigenvalues of $M$ can be directly determined on the lattice \cite{s8,s13,s9}. 

A diagnostic for confinement can be obtained by making the replacement
\begin{equation}
\langle\rho^a(x)\rho^c(z)\rangle\rightarrow \frac{Q^2}{V}\delta(x-z)\delta_{ac}
.\end{equation}
This will yield the self-energy of a point color charge $Q$. This Coulomb self-energy is
\begin{equation}
E_{\rm self}=\frac{Q^2}{2V}\sum_n \frac{F_{n}}{\lambda_n^2}
,\end{equation}
where $F_n=F_{nn}$.
In the large volume limit the sum can be replaced by an integral upon introducing a density of states  \cite{s8}, 
\begin{equation}
E_{\rm self}=\frac{1}{2}Q^2\int d\lambda \rho(\lambda)\frac{F(\lambda)}{\lambda^2}
\label{e10}.\end{equation}

In the weakly interacting limit the eigenstates $\phi_n^a(x)$ are plane waves and $\rho(\lambda)=\sqrt{\lambda}/4\pi^2$ and $F(\lambda)=\lambda$, which means that $E_{\rm self}$ has an ultraviolet divergence but no infrared divergence. But confinement requires that $E_{\rm self}$ does exhibit an infrared divergence, and so we know that QCD must produce an infrared enhancement of $\rho(\lambda)$ and/or $F(\lambda)$. If we write
\begin{equation}
\rho(\lambda)\approx \frac{\sqrt{\lambda}}{4\pi^2}\left(\frac{\Lambda^2}{\lambda}\right)^a
,\quad\quad\quad
F(\lambda)\approx \lambda\left(\frac{\Lambda^2}{\lambda}\right)^b
,\label{e7}\end{equation}
then we see that the existence of confinement implies that $a+b>1/2$ \cite{s8}. We are using $\Lambda\equiv\Lambda_{\rm QCD}$ to represent the scale below which confinement physics enters and in the following it will also represent the mass gap of the theory.

Let us return to (\ref{e1}) and consider the vacuum energy density $\rho_{\rm vac}=H_{\rm coul}/V$. Now we need the quantity $\langle\rho^a(x)\rho^c(z)\rangle$. The nonperturbative contribution to this quantity is easier to determine than $M^{-1}$ since it is almost completely determined by the mass gap. The fields in $\rho^a(x)$ from (\ref{e6}) are the transverse gluons and the light quarks (of number $n_q$). The quarks develop a dynamical mass (we ignore current quark masses) and the transverse gluons have an effective mass as mentioned above. But the fluctuations of these fields do not in turn produce a mass for the Coulomb potential field. From this and dimensional analysis, the lowest order term in the derivative expansion of $\langle\rho^a(x)\rho^c(z)\rangle$ should take the form
\begin{equation}
\langle\rho^a(x)\rho^c(z)\rangle\propto \Lambda\nabla^2_x\delta(x-z)\delta_{ac}
.\end{equation}
This argument does not determine the sign.

To see more clearly where such a result comes from it is instructive to consider the one loop contribution evaluated in momentum space,
\begin{equation}
\langle\rho^a(x)\rho^c(z)\rangle\approx\delta_{ac}\int \frac{d^3p}{(2\pi)^3} e^{i(x-z)\cdot p}\left(\frac{\alpha_s}{2\pi}(1+\frac{2}{3}n_q)p^2\int dk \left[\frac{ k^2}{\omega_k^2}-1\right]\right)
.\label{e2}\end{equation}
We have kept only the lowest order term of an expansion in $p^2$. $\Lambda$ now appears in $\omega_k^2=k^2+\Lambda^2$ as an approximation to the mass of both the quarks and the transverse gluons. We have performed a subtraction by the zero mass integral to isolate the contribution that vanishes with the mass gap.

The framework here is non-covariant old fashioned perturbation theory (OFPT). The result in (\ref{e2}) is closely related to the transverse gluon and quark contribution to the usual vacuum polarization \cite{s10}. That contribution occurs at second order in OFPT where the quantity $\sum_n\langle\rho^a(x)|n\rangle\langle n|\rho^c(z)\rangle/(E_0-E_n)$ appears. The extra energy denominator turns the linear divergence appearing before subtraction in (\ref{e2}) into the standard log divergence. The transverse gluons and quarks give screening contributions to the vacuum polarization, and so the sign of our result has the same origin. Meanwhile the anti-screening contribution to the vacuum polarization enters at first order in OFPT and it comes from the perturbative expansion of $M^{-1}$ \cite{s10}.

The result of the $k$ integral in (\ref{e2}) is $-\Lambda\pi/2$, and along with the approximations $\alpha_s=1$ and $n_q=3$ we arrive at
\begin{equation}
\langle\rho^a(x)\rho^c(z)\rangle\approx\frac{3}{4}\Lambda\nabla^2_x\delta(x-z)\delta_{ac}
.\label{e12}\end{equation}
While this perturbative derivation of a nonperturbative effect is no doubt an oversimplification, we shall continue to use (\ref{e12}) in the following. The calculation clarifies the role of both the mass and the screening effects of the transverse gluons and quarks.

Inserting this result back into $\rho_{\rm vac}=H_{\rm coul}/V$ from (\ref{e1}) gives
\begin{equation}
\rho_{\rm vac}\approx-\frac{3\Lambda}{8 V}\sum_{mn} \frac{F_{nm}F_{mn}}{\lambda_m\lambda_n}
.\end{equation}
According to a lattice study \cite{s9} the off-diagonal elements are small, so $F_{mn}\approx F_n\delta_{mn}$. If we use this then
\begin{eqnarray}
\rho_{\rm vac}&\approx&-\frac{3\Lambda}{8 V}\sum_n\frac{F_n^2}{\lambda_n^2}\\
&\approx&-\frac{3\Lambda}{8 }\int d\lambda \rho(\lambda)\frac{F(\lambda)^2}{\lambda^2}\\
&\approx&-\frac{3\Lambda}{32\pi^2}\int d\lambda\sqrt{\lambda}\left(\frac{\Lambda^2}{\lambda}\right)^{(a+2b)}
\label{e11}\end{eqnarray}

We are interested in the infrared behavior of this integral. Suppose there is some large length scale $L$ beyond which the physics changes and the infrared enhancement effects are reduced (see below). Corresponding to $L$ would be a minimum $\lambda_{\rm min}$ below which the integration in (\ref{e11}) would change. The modified integrand is reduced for example if it scales with a higher power of $\lambda$ for $\lambda<\lambda_{\rm min}$. In any case we shall take the shift in the vacuum energy density due to this change to be of order the negative of the 0 to $\lambda_{\rm min}$ part of the integration in (\ref{e11}).

To go further we need to specify how $\lambda_{\rm min}$ scales with $L$, 
\begin{equation}
\lambda_{\rm min}\approx \frac{1}{L^2}\frac{1}{(L\Lambda)^c}
.\label{e8}\end{equation}
We shall interpret $\lambda_{\rm min}$ as the minimum nonzero eigenvalue of $M$ that exists in a finite volume $L^3$. The resulting finite shift in vacuum energy density is
\begin{eqnarray}
\Delta\rho_{\rm vac}&\approx& \frac{3}{32\pi^2}\Lambda^{2a+4b+1}\lambda_{\rm min}^{\frac{3}{2}-a-2b}\nonumber\\&\approx&\frac{3}{32\pi^2}\Lambda^{4-d}(\frac{1}{L})^d
\label{e5}\end{eqnarray}
where
\begin{equation}
d=(2+c)(\frac{3}{2}-a-2b)
.\label{e9}\end{equation}
In the weak interaction limit $a=b=c=0$, and $d=3$. Although it doesn't make much sense to consider this limit without also setting the mass gap to zero, one can see the similarity between $d=3$ and the naive estimate in (\ref{e13}).

$\Delta\rho_{\rm vac}$ is enhanced as $d$ decreases, and we already see that the bound $a+b>1/2$ from confinement will tend to do this. To do better we shall use the results of an $SU(3)$ lattice study \cite{s9}.\footnote{An earlier lattice study \cite{s8} for $SU(2)$ found $a\approx.25$ and $b\approx.62$.} There the eigenvalue equation
\begin{equation}
M\phi_n=\lambda_n\phi_n
\end{equation}
was solved on a lattice where $M$ is a $(8\cdot L^3)\times(8\cdot L^3)$ matrix. The volume dependence of the lowest eigenvalues was studied and it was found that $c\approx0.5$. To obtain $a$ and $b$ we notice that the quantities $\lambda_n$ and $F_n/\lambda_n^2$ were obtained for the lowest 292 nontrivial eigenvalues. In Fig.~(1) we display both $n$ and $F_n/\lambda_n^2$ vs. $\lambda_n$ on a log-log plot for $L=24$.\footnote{The authors of \cite{s9} kindly provided me with their raw data for the $L=12$, 16 and 24 cases.} Since $n\propto\int^\lambda d\lambda'\rho(\lambda')$, the slopes of these two curves can be used to determine $a$ and $b$. In Fig.~(2) we show the residuals (actual values divided by fit values) of the linear fits, where the least squares weights are determined by the statistical errors.\footnote{Very similar results follow from an unweighted fit where the lowest 27 eigenvalues are dropped.} We also perform fits for the $L=12$ and 16 data. The structure observed in these residuals, the oscillations and the increasing departures for the smallest $\lambda_n$'s, are due to finite volume effects. The observed oscillations can be expected to continue for some range of $n>292$ as well. This makes the extracted slopes somewhat sensitive to the range of $n$ used, and so these results for $n\le292$ should only be considered as indicative.
\vspace{-0ex}\begin{center}\includegraphics[scale=0.5]{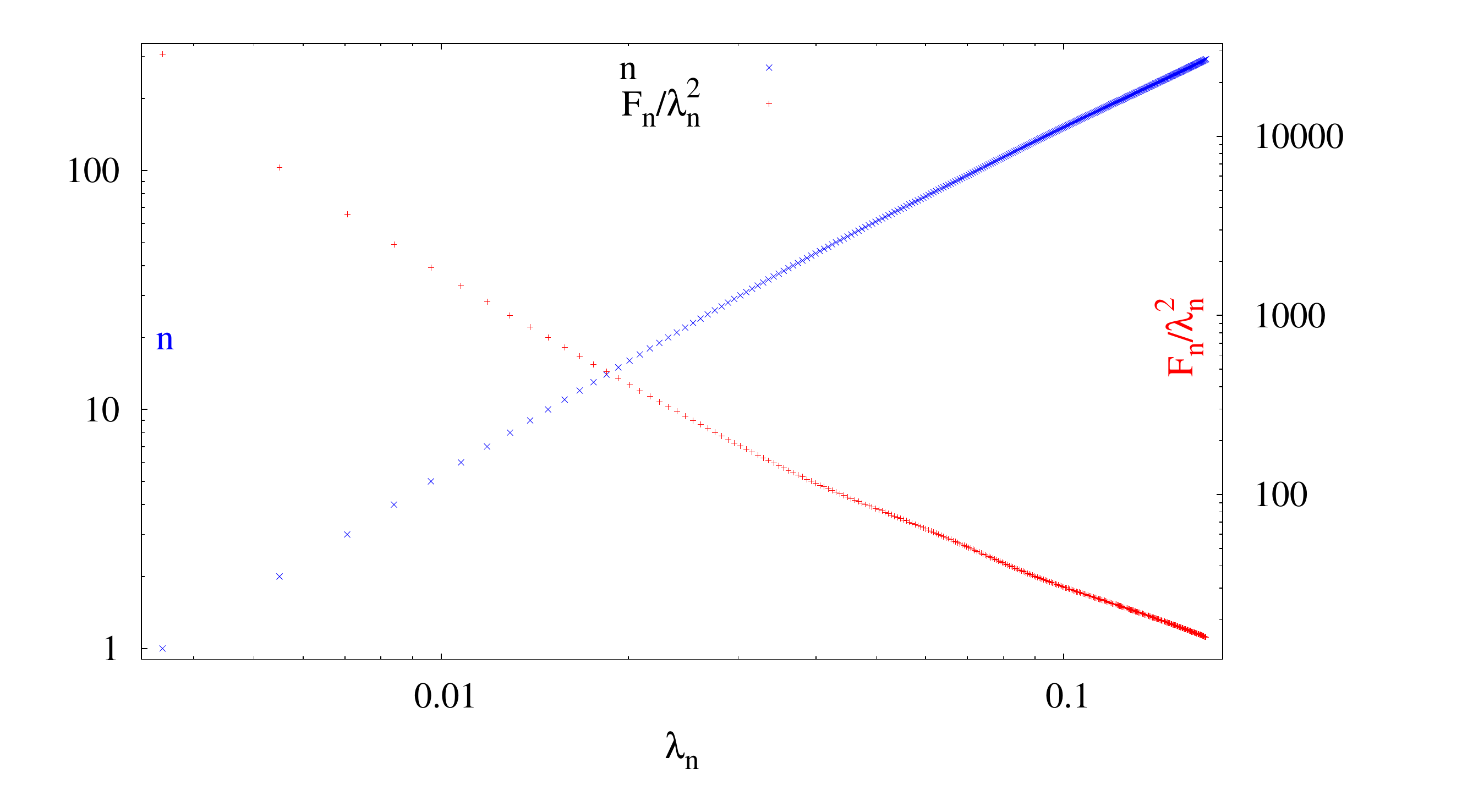}\end{center}
\vspace{-4ex}\noindent Figure 1: The quantities $n$ and $F_n/\lambda_n^2$ are plotted vs. $\lambda_n$ for the lowest 292 eigenvalues for $L=24$, from a lattice study \cite{s9}.
\begin{center}\includegraphics[scale=0.5]{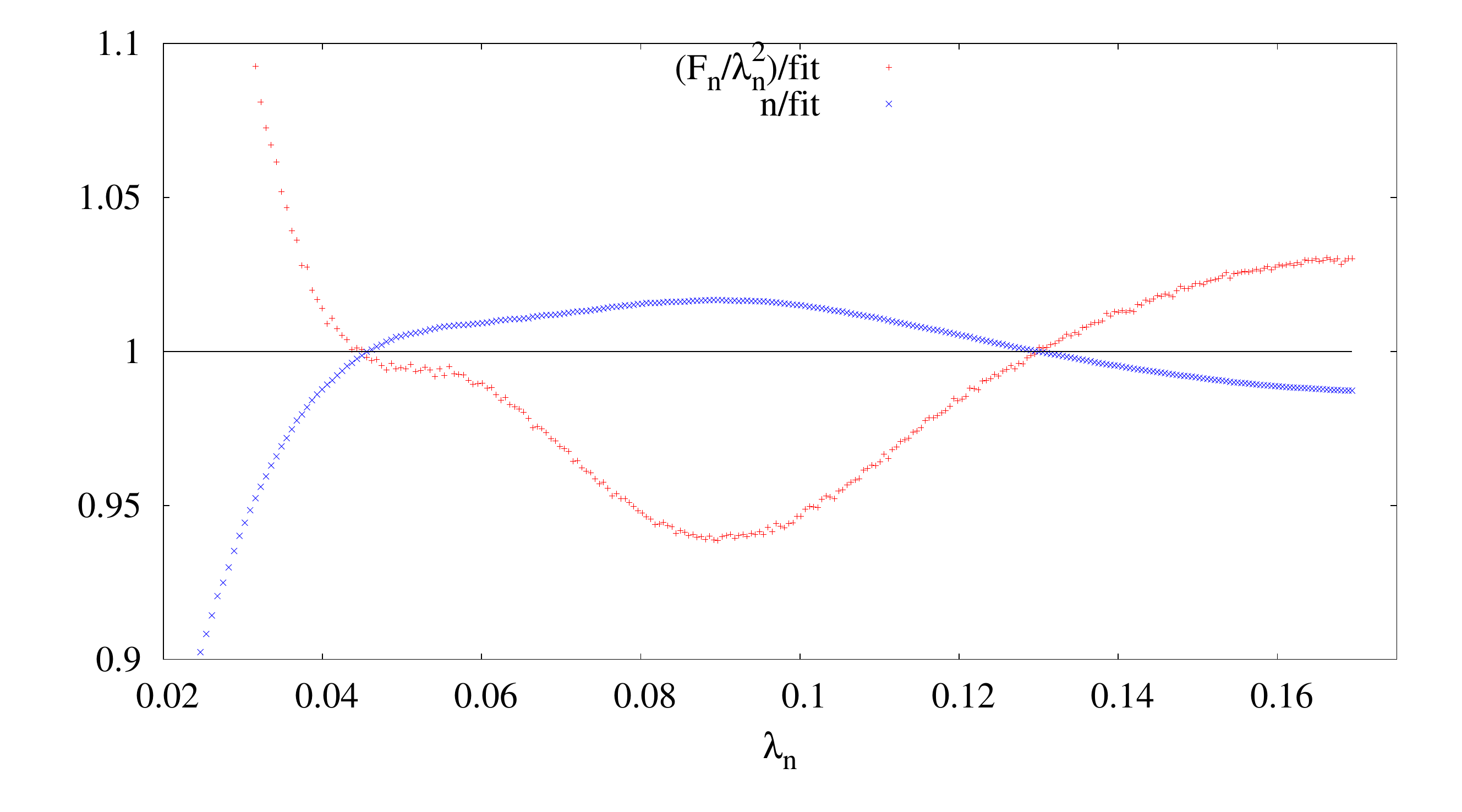}\end{center}
\vspace{-4ex}\noindent Figure 2: The residuals of the linear fits  (actual divided by fit values).
\vspace{2ex}

We display our fits for $a$ or $b$ in Table 1, and the $d$ values assume $c=0.5$. We see that the $L=12$ and 16 results do not satisfy the constraint $a+b>1/2$. Although the results in Table 1 have uncertainties due to the finite volume effects, the trend of decreasing $d$ for increasing $L$ is clear. Presumably larger lattice sizes are needed to find the limiting value of $d$. In the cosmological context a value of $d$ close to unity may give a vacuum energy of the correct magnitude. The results in Table 1 are still consistent with this, and they clearly motivate more refined lattice determinations of the scaling exponents. It is interesting that cosmology can provide at least a lower bound on a scaling parameter measured in lattice gauge theory simulations.
\begin{center}
\begin{tabular}{|c|c|c|c|c|}\hline $L$ & $a$ & $b$ & $a+b$ & $d$ \\\hline 12 & .094 & .238 & .33 & 2.34 \\\hline 16 & .146 & .285 & .43 & 1.96 \\\hline 24 & .205 & .381 & .59 & 1.33 \\\hline \end{tabular}
\end{center}
\vspace{-0ex}\noindent Table 1: Fit values for the scaling exponents appearing in (\ref{e7}), (\ref{e8}) and (\ref{e9}).

\section{The cosmology}
In the last section we obtained the shift in the QCD vacuum energy density $\Delta\rho_{\rm vac}$ due to some change of physics occurring beyond a length scale $L$. This introduced a dependence of $\Delta\rho_{\rm vac}$ on $1/L$, and now in the cosmological context we consider the replacement of $1/L$ by $H(t)$. Hubble expansion changes the physics on and beyond Hubble scales in the sense that the vacuum fields are not able to attain the true Minkowskian vacuum state of QCD. It appears reasonable that the Hubble scale is the relevant scale, since points separated by more than the Hubble distance are receding from each other faster than the speed of light. This is true whether or not the universe is accelerating. Certainly the Minkowski description of the QCD vacuum would break down on these scales, since the interactions and correlations on larger scales would be disrupted. In particular it is fair to assume that the enhancement that these long wavelength modes enjoy in the true QCD vacuum no longer occurs. This is all that is needed to support our discussion below (\ref{e11}).

So assuming that this residual energy density exists, the question is how it affects the evolution of the universe. This energy density cannot be described by some local term in an action, constructed to reproduce the $H(t)$ dependence, since such a term would respond to all wavelengths. Implementing an effect only on wavelengths $\gtrsim H(t)^{-1}$ through the action would be highly nonlocal. Nevertheless there should be an effective energy-momentum tensor $T^D_{\mu\nu}$, depending on $H(t)$, that can serve to describe the back-reaction effects on the cosmological evolution. There is also no reason for such a $T^D_{\mu\nu}$ to be proportional to $g_{\mu\nu}$, since over the relevant distance scales the Lorentz violating nature of the preferred cosmological frame is fully manifest. Nevertheless, in the cosmological rest frame, $T^D_{\mu\nu}$ should preserve spatial isotropy and thus it should have a perfect fluid form in terms of the energy density $\rho_D$ and an effective pressure $p_D$.

Next is the question of whether the dark energy exchanges energy with the normal matter and radiation content of the universe. Since the typical wavelengths of these normal components are vastly smaller than the dark energy wavelengths, we do not expect energy exchange. If there are fields with modes of Hubble wavelengths being excited for other reasons, then we can simply redefine $T^D_{\mu\nu}$ to include these contributions (assuming that they give subdominant contributions). We shall thus suppose that this full dark energy $T^D_{\mu\nu}$ does not interact with normal matter and radiation. In this case we must have conservation, $\partial^\mu T^D_{\mu\nu}=0$. This will relate the effective $p_D$ to $\rho_D$, with both quantities vanishing with $H$.

Let us consider the evolution of quantities in terms of the scale parameter $a$ rather than time. The dark energy density is $\rho_D(a)=CH(a)^{d}$ with our estimate from (\ref{e5}) being $C\approx \Lambda_{\rm QCD}^{4-d}3/32\pi^2$. The Friedmann equation becomes 
\begin{equation}
\rho(a)+CH(a)^{d} =\frac{3}{8\pi G}H(a)^{2}, \quad\quad 0\le d<2
.\label{e3}\end{equation}
In the early universe the normal energy density $\rho(a)$ starts off larger than $\rho_D(a)$, but it decreases faster as $a\rightarrow\infty$. In this limit $H(a)$ is driven towards a  constant
\begin{equation}
H(a\rightarrow\infty)=\left(\frac{8\pi GC}{3}\right)^\frac{1}{2-d}
\end{equation}
from above. Thus the universe is driven to a de Sitter expansion.

When we insert this limiting value back into $\rho_D(a)=CH(a)^{d}$ and require that this corresponds to the observed value of dark energy $(2.4\times10^{-3}$ eV)$^4$ (and correct for the fact that the latter is the present value rather than the limiting value), we obtain an estimate for $d$. If we take $\Lambda_{\rm QCD}=300$ MeV then $d\approx 1.026$. Meanwhile $d=1$ and 1.05 correspond to $\Lambda_{\rm QCD}=130$ and 650 MeV respectively. And one should keep in mind uncertainties in the estimate of $C$.

It is useful to keep $d$ in the formulas (some similar results for $d=1$ have appeared recently in \cite{s14}). We can assume that $\rho(a)$ describes only pressureless matter, so $\rho(a)=\rho_0/a^3$, and we take $a=1$ at the present time. From (\ref{e3}) we can obtain
\begin{equation}
H'(a)=\frac{H(a)}{a}\frac{3\Omega_m(a)}{d-2-d\Omega_m(a)}
,\label{e4}\end{equation}
where $\Omega_m(a)$ is the matter fraction of the total energy density. Using this we can determine the inflection point where $\ddot{a}(t)=0$, or in other words when $(aH(a))'=0$. It occurs when
\begin{equation}
\Omega_m(a_{\rm infl})=\frac{2-d}{3-d}
.\end{equation}
Another dimensionless quantity of interest at the present time is $t_0H_0$. $t_0$ can be obtained through a numerical integration of $\dot{a}(t)=a(t)H(a(t))$.

Values of the inflection point red shift, $z_{\rm infl}=1/a_{\rm infl}-1$, and $t_0H_0$ can be obtained as a function of the present value of $\Omega_m\equiv\Omega_m(1)$. In Table 2 we compare the $d=1.05$ and 1 values with the $d=0$ values of a cosmological constant cosmology (note the different $\Omega_m$ values).  We see that $z_{\rm infl}$ and $t_0H_0$ for the two cosmologies become more similar if $\Omega_m$ is reduced by about 0.04 or 0.05 when going from $d=0$ to $d=1$ or 1.05.

\vspace{1ex}
\begin{center}\begin{tabular}{|c|c|c|c|}\hline $\Omega_m$ & $z_{\rm infl}$ & $t_0H_0$ & $w_D$
\\\hline .20 & .838 & .998 & $-$.819 
\\\hline .21 & .793 & .986 & $-$.812
\\\hline .22 & .749 & .975 & $-$.804
\\\hline .23 & .708 & .964 & $-$.797
\\\hline .24 & .669 & .954 & $-$.790 \\\hline\multicolumn{4}{c}{$d=1.05$} \end{tabular}
\quad\begin{tabular}{|c|c|c|c|}\hline $\Omega_m$ & $z_{\rm infl}$ & $t_0H_0$ & $w_D$
\\\hline .20 & .857 & 1.004 & $-$.833 
\\\hline .21 & .811 & .992 & $-$.826
\\\hline .22 & .768 & .980 & $-$.820
\\\hline .23 & .728 & .970 & $-$.813
\\\hline .24 & .688 & .959 & $-$.806 \\\hline\multicolumn{4}{c}{$d=1$} \end{tabular}
\quad\begin{tabular}{|c|c|c|c|}\hline $\Omega_m$ & $z_{\rm infl}$ & $t_0H_0$ & $w_D$ 
\\\hline .24 & .850 & 1.025 & $-1$ 
\\\hline .25 &.817 & 1.014 & $-1$ 
\\\hline .26 & .786 & 1.003 & $-1$ 
\\\hline .27 & .756 & .993 & $-1$ 
\\\hline .28 & .726 & .983 & $-1$ \\\hline\multicolumn{4}{c}{$d=0$} \end{tabular}\\
Table 2
\end{center}
\begin{center}\includegraphics[scale=0.5]{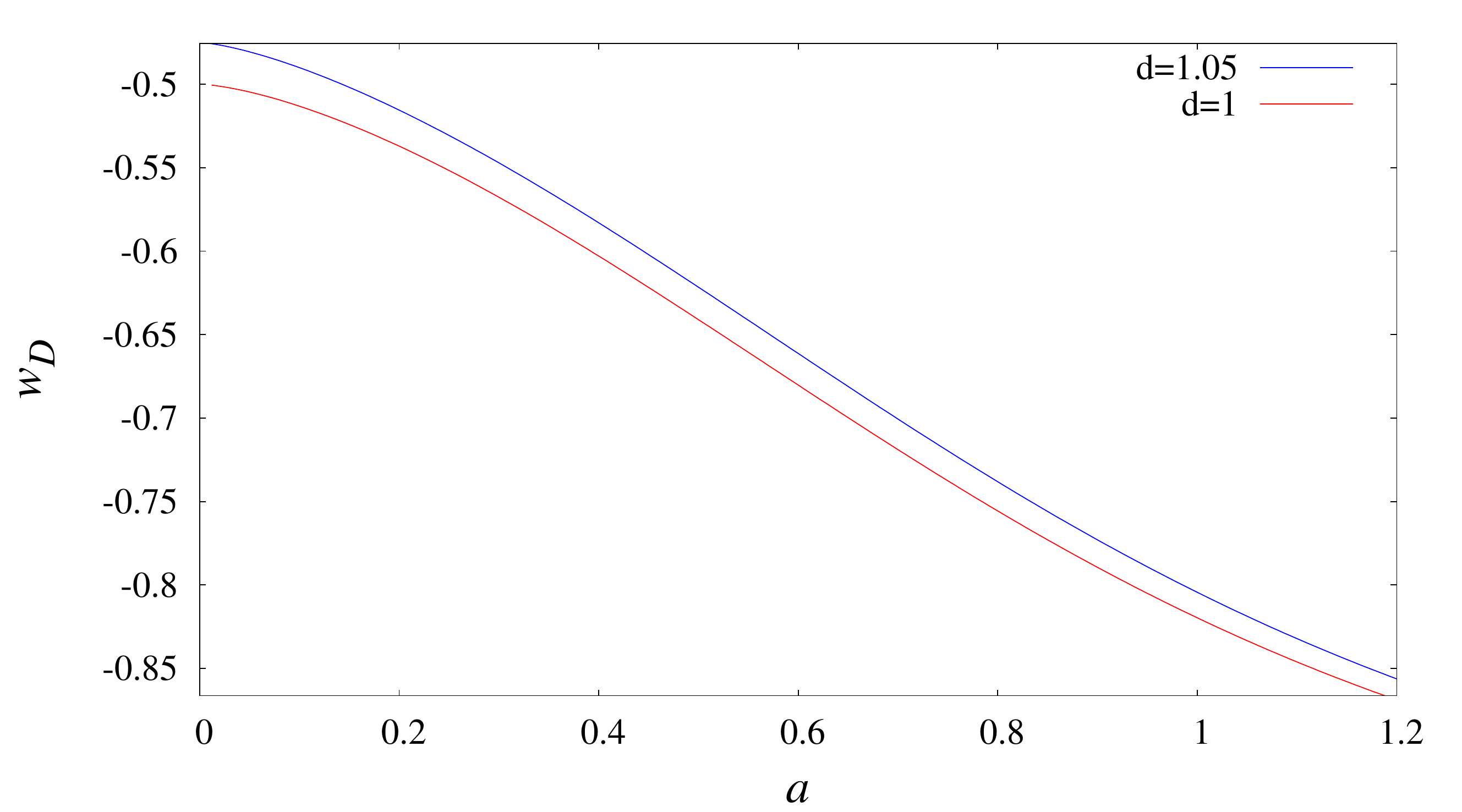}\end{center}
\vspace{-1ex}\noindent Figure 3: The effective equation of state parameter for dark energy when $\Omega_m=0.22$. $a=1$ is the present.
\vspace{1ex}

The conservation equation $\rho_D'+3(1+w_D)\rho_D/a=0$ and (\ref{e4}) determine the effective equation of state parameter,
\begin{eqnarray}
w_D(a)&=&-1-\frac{d}{3}\frac{aH'(a)}{H(a)}\nonumber\\
&=&\frac{d-2}{2+d(\Omega_m(a)-1)}
.\end{eqnarray}
Thus $w_D(a)$ evolves from a value of $(d-2)/2$ when dark energy is negligible to a value of $-1$ when it comes to dominate. At the inflection point $w_D(a_{\rm infl})=(d-3)/3$. The present values $w_D\equiv w_D(1)$ are given in the Table 2 while examples of the evolution of $w_D(a)$ are shown in Fig.~(3).\footnote{In the case $d=1$ we have the more explicit result \[\Omega_m(a)/4a={{\left( \sqrt{4\,a+\frac{{a}^{4}\,{\left( 1-\Omega_m\right) }^{2}}{\Omega_m}}+\frac{{a}^{2}\,\left( 1-\Omega_m\right) }{\sqrt{\Omega_m}}\right) }^{-2}}.\]} And finally the form of the basic equation that describes structure formation in normal matter $\ddot{\delta}+2H\dot{\delta}-\frac{3}{2}\Omega_m H^2\delta=0$ does not change, where again the dark energy does not participate at sub-Hubble wavelengths.

For the cosmology then, the main observation is that the present values $\Omega_m$ and $w_D$ are both reduced as compared to the cosmological constant cosmology. These reductions appear to be consistent with present data. In particular we highlight a recent analysis \cite{s12} of a combination of datasets that have a similar intermediate redshift, with the goal of reducing systematic uncertainties. The values as quoted there are $\Omega_m=0.22^{+0.09}_{-0.08}$ and $w_D=-0.81^{+0.16}_{-0.18}(\mathrm{stat}){\pm0.15}(\mathrm{sys})$. The analysis of more recent and future datasets, as is already underway, will presumably go further at distinguishing a QCD generated dark energy from a standard cosmological constant.

An interesting implication of these considerations is that they constrain the existence of confining gauge theories with mass scales higher than $\Lambda_{\rm QCD}$. Their contribution to dark energy could be too large unless their corresponding $d$ parameter is also larger. Conversely if such a contribution was the dark energy, then the larger value of $d$ would give a cosmology further from the cosmological constant cosmology. For example if $\Lambda=1$ TeV then $d\approx1.245$, which gives $w_D=-0.73$ for $\Omega_m=0.22$.

To sum up, QCD is confining and has a mass gap. Nevertheless the long wavelength structure of the QCD vacuum differs markedly from that of free fields, either massive or massless, as can be seen in its response to the introduction of a color charge. We have argued that this implies significantly enhanced long wavelength contributions to the vacuum energy as well. But the true Minkowskian QCD vacuum presumably cannot be attained in an expanding universe, and this leads to a residual energy density that can be discussed independently of the cosmological constant problem. The true vacuum is approached as the universe expands, but eventually the residual energy dominates and the universe becomes stuck in a de Sitter expansion. The Minkowski vacuum is never attained. The question of whether the QCD residual energy is of the right size to be the present dark energy remains to be seen. It is interesting that cosmological observations and lattice studies can both come to bear on this question.

\section*{Acknowledgments}
I thank Y.~Nakagawa for correspondence and data relating to reference \cite{s9}. I also thank F.~Klinkhamer for sparking my interest in this topic. This work was supported in part by the Natural Science and Engineering Research Council of Canada.


\begin{thebibliography}{11}
\bibitem{s1} R.~Schutzhold, Phys.~Rev.~Lett.~89 (2002) 081302, gr-qc/0204018.
\bibitem{s2} F.R.~Klinkhamer, G.E.~Volovik, Phys.~Rev.~D79:063527, 2009, arXiv:0811.4347; F.R.~Klinkhamer, Phys.~Rev.~D81:043006, 2010,  arXiv:0904.3276; F.R.~Klinkhamer, ``QCD-scale modified-gravity universe'', arXiv:1005.2885.
\bibitem{s3} E.C.~Thomas, F.R.~Urban, A.R.~Zhitnitsky, JHEP 0908:043,2009, arXiv:0904.3779; F.R.~Urban, A.R.~Zhitnitsky, Phys.~Lett.~B688:9, 2010, arXiv:0906.2162;  JCAP0909:018, 2009, arXiv:0906.3546; Nucl.Phys.B835:135, 2010, arXiv:0909.2684.
\bibitem{s6} F.R.~Urban, A.R.~Zhitnitsky, Phys.~Rev.~D80:063001,2009, arXiv:0906.2165; A.R.~Zhitnitsky, Phys.~Rev.~D82, 103520 (2010),  arXiv:1004.2040
\bibitem{s7} N.~Ohta, Dark Energy and QCD Ghost, arXiv:1010.1339.
\bibitem{s5} V.~N.~Gribov, Nucl.~Phys.~B139 (1978) 1.
\bibitem{s15} D.~Zwanziger, Prog.~Theor.~Phys.~Suppl.~131, 233 (1998), hep-th/9802180;  A.~Cucchieri and D.~Zwanziger, Phys.~Rev.~D65, 014001 (2001), hep-lat/0008026; A.~Cucchieri, AIP Conf.~Proc.~892, 22 (2007), hep-lat/0612004.
\bibitem{s4} G.~Burgio, M.~Quandt and H.~Reinhardt, Phys.~Rev.~Lett.~102, 032002 (2009), arXiv:0807.3291; PoS CONFINEMENT8, 051 (2008), arXiv:0812.3786.
\bibitem{s16} A.~Maas, Gauges, propagators, and physics, arXiv:1011.5409.
\bibitem{s8} J.~Greensite, S.~Olejnik, D.~Zwanziger, JHEP 0505:070, 2005, hep-lat/0407032.
\bibitem{s9} Y.~Nakagawa, A.~Nakamura, T.~Saito, H.~Toki, Phys.~Rev.~D81:054509, 2010, arXiv:1003.4792.
\bibitem{s13} Y.~Nakagawa, A.~Nakamura, T.~Saito and H.~Toki, Phys.~Rev.~D75, 014508 (2007), hep-lat/0702002.
\bibitem{s10} S.~Drell,  in: G.~Feinberg, A.W.~Sunyar, J.~Weneser (Eds.), Transactions of the New York Academy of Sciences, Series II, Vol.~40, New York Acad. Sci., New York 1980, SLAC-PUB-2694 (reprint), and references therein.
\bibitem{s14} R-G.~Cai, Q.~Su, Z-L.~Tuo, H-B.~Zhang, Notes on Ghost Dark Energy, arXiv:1011.3212.
\bibitem{s12} H.~Lampeitl et.al., Mon.~Not.~Roy.~Astron.~Soc.~401:2331, 2009, arXiv:0910.2193.



\end{thebibliography}
\end{document}